\documentclass[prb,twocolumn,showpacs,preprintnumbers,amsmath,amssymb]{revtex4}
\usepackage{dcolumn}
\usepackage{bm}
\usepackage{amsmath,graphicx}
\usepackage{color}

\newcommand{\ep}{\varepsilon}
\newcommand{\pa}{\partial}

\begin{document}
\title{Diverse corrugation pattern in radially shrinking carbon nanotubes}
\author{Hiroyuki Shima}
\email{shima@eng.hokudai.ac.jp}
\affiliation{
Division of Applied Physics, 
Faculty of Engineering,
Hokkaido University, Sapporo, Hokkaido 060-8628, Japan}
\affiliation{Department of Applied Mathematics 3, LaC\`aN, Universitat Polit\`ecnica de Catalunya, Barcelona 08034, Spain}

\author{Motohiro Sato}
\affiliation{Division of Socio-Environmental Engineering, Faculty of Engineering, Hokkaido University, Sapporo, Hokkaido 060-8628 Japan}

\author{Kohtaroh Iiboshi}
\affiliation{Division of Socio-Environmental Engineering, Graduate School of Engineering, Hokkaido University, Sapporo, Hokkaido 060-8628 Japan}

\author{Susanta Ghosh}
\affiliation{Department of Applied Mathematics 3, LaC\`aN, Universitat Polit\`ecnica de Catalunya, Barcelona 08034, Spain}

\author{Marino Arroyo}
\affiliation{Department of Applied Mathematics 3, LaC\`aN, Universitat Polit\`ecnica de Catalunya, Barcelona 08034, Spain}

\date{\today}

%
%

\begin{abstract}
Stable cross-sections of multi-walled carbon nanotubes subjected to electron-beam irradiation are investigated in the realm of the continuum mechanics approximation. The self-healing nature of sp$^2$ graphitic sheets implies that selective irradiation of the outermost walls causes their radial shrinkage with the remaining inner walls undamaged. The shrinking walls exert high pressure on the interior part of nanotubes, yielding a wide variety of radial corrugation patterns ({\it i.e.,} circumferentially wrinkling structures) in the cross section. All corrugation patterns can be classified into two deformation phases for which the corrugation amplitudes of the innermost wall differ significantly.
\end{abstract}




%
%
%

\maketitle

\section{Introduction}

Carbon nanotubes exhibit remarkable flexibility when subjected to
cross-sectional deformation.
\cite{Despres1995,Iijima1996,Palaci2005} Such flexibility
has been evidenced by spectroscopy and diffraction measurements in
which hydrostatic pressure on the order of a few GPa
caused flattening and polygonization in the cross section.
\cite{Peters2000,J_Tang2000,Elliott2004}
Molecular-dynamics (MD)
\cite{Tangney2005,Gadagkar2006,Zhang2006} and
density-functional-theory-based
simulations\cite{Reich2002,Chan2003}
revealed more details about the mechanism of radial deformation
under pressure. Radial deformation can also be caused by a
nanoindentation \cite{Barboza_2008} under which the
elastic response of single-walled nanotubes (SWNTs) is determined
by a universal law that depends on the tube radius $r$
and applied load.\cite{Barboza_2009} 
As the radial deformation of
nanotubes is strongly correlated with their electronic
\cite{Barboza_2008,Park1999,Mazzoni2000,Gomez2006}
and optical \cite{Lebedkin2006} properties, its
clarification should be important from the viewpoint of the use of
nanotubes as components of nanoscale devices.

The cross section of an isolated SWNT exhibits a
circular-to-elliptic transition at a critical hydrostatic
pressure $p_{\rm c}$. The value of $p_c$ decreases with $r$ as
\cite{Yakobson1996,DYSun2004} $p_{\rm c} \propto r^{-3}$, which
agrees with the conclusion of the classical continuum theory.
\cite{Brush_1975} This scenario, however, fails in the case of
multi-walled nanotubes (MWNTs). It has been shown that
MWNTs consisting of several tens of concentric walls undergo a
novel cross-sectional deformation, called radial corrugation,
\cite{NTN} in which the outer walls exhibit wavy structures along
the circumferential direction. 
The radial corrugation originates from the multilayered
nature, {\it i.e.,} the competing effects between the mechanical
instability of outer walls with large
radii and the radial rigidity of inner walls with small radii.

The corrugation behavior of a MWNT is believed to change if it
occurs in an embedding elastic medium. Suppose that a MWNT is
embedded in a large elastic medium that is in complete contact
with its outermost wall. A uniform shrinkage of the medium exerts
an external force on the outermost wall, as similar to the case of
hydrostatic pressure. 
But a difference arises when the medium possesses moderate stiffness,
since the energy required to deform the medium should be responsible
for determining the stable corrugation pattern.
In fact, when a thin {\it flat} plate is 
embedded in an elastic medium and comressed,
it shows a uni-directional corrugation whose profile is strongly
dependent on the medium's stiffness.\cite{Allen_1969,Volynskii_2000}
This fact implies an uncovered corrugation mechanism peculiar to 
embedded MWNTs,
{it i.e.,} dinstinct from the hydrostatic pressure cases,
which imposes new challenges.

In this article, we demonstrate that the presence of a surrounding
elastic medium triggers diverse variations of corrugation modes
that cannot be observed in MWNTs under hydrostatic pressure.
These diverse modes are found to be grouped into
two corrugation phases that exhibit a significant difference in
the corrugation amplitude of the innermost wall of the MWNT. These
findings will help to tune the core-tube geometry of MWNTs, thus
providing useful information for developing nanofluidic
\cite{Majumder2005,Noy2007,Whitby2007} or
nanoelectrochemical \cite{Frackowiak2001} devices
whose performance depends on the geometry of the inner hollow
cavity of nanotubes.

\section{Methodology}

\subsection{Self-healing nature of graphite sheets}

The current work has been inspired by the discovery of the
self-healing properties of carbon nanostructures under
electron-beam irradiation.\cite{Banhart1996,Ajayan1998,Banhart2005,LSun2006,Krasheninnikov2007}
Banhart {\it et al.} \cite{Banhart1996} found that carbon onions,
assemblies of concentric spherical carbon shells, shrink toward
smaller onion structures when irradiated. This shrinkage is
attributed to the knock-on collision of carbon atoms from
constituent shells; the resulting vacancies are healed by
annealing reconstruction \cite{Ding2007}
that shrinks the system with reduced
intershell spacings. As a result, extreme pressure arises in the
core of spherical carbon shells. Using a similar irradiation
technique, Sun {\it et al.} \cite{LSun2006} synthesized MWNTs with
reduced interwall spacing and revealed high pressures of several
tens GPa within the hollow cavity. 
It is thus expected that self-contraction of {\it only} outermost walls of a MWNT
(possibly realized by finely-focused beam)
will provide the setting for radial corrugation of the interior part
of the MWNT.

It should be borne in mind that in the carbon onion experiments,
diamond nucleation occurs at the center of the onion
as the induced pressure is too high for the original sp$^2$ bonds to persist.\cite{Banhart1996}
To make the current work convincing, therefore, 
it is important to examine whether the critical pressure $p_c$
for the corrugation is low enough to prevent the breaking
of sp$^2$ bonds within the inner walls.
We will revisit this issue in Section III B.
%

\begin{figure}[ttt]
\includegraphics[width=8.8cm]{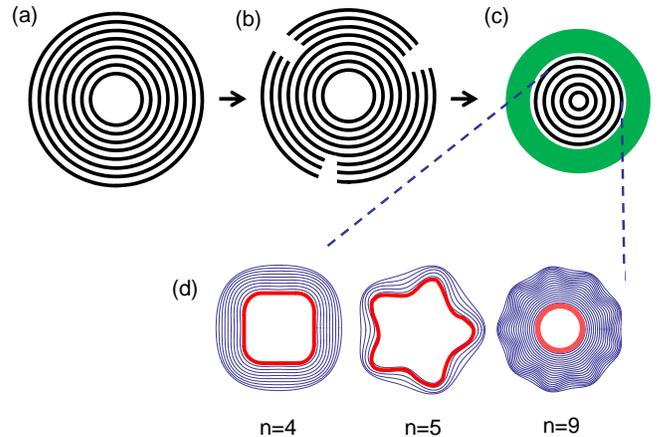}
\caption{(color online) Illustration of radial shrinking hypothesis of MWNTs. (a)
An $(M+N)$-walled nanotube is subjected to electron-beam
irradiation. (b) Irradiation with fine resolution kicks off a
portion of carbon atoms located within $M (\gg N)$ outermost
walls. (c) The eroded $M$ walls are healed and radially contracting,
as a result of which extreme pressure is exerted 
on the $N$ innermost walls. (d) Typical corrugation modes of the
pressed $N$-walled nanotubes are shown with the mode index $n$.} \label{fig_1}
\end{figure}

%
\subsection{Continuum model of radially shrinking MWNT}

When treating many-walled nanotubes,
atomistic simulations are realistic and accurate
but they demand huge computational cost in general.
Thus we used a simplified model based on the
continuum approximation.
\cite{Yakobson1996, Ru,Pantano2004,Huang} 
First, we map an MWNT onto concentric cylindrical shells endowed with
\cite{Kudin_PRB2001}
the in-plane stiffness $C = 345$ nN/nm,
the flexural rigidity $D = 0.238$ nN$\cdot$nm,
and Poisson's ratio $\nu=0.149$,
in which the inter-shell separation is defined by 0.3415 nm.
Figures \ref{fig_1}(a)-(c) illustrate the shrinking process
considered in this paper. We assume an $(M+N)$-walled nanotube
under irradiation that kicks off a portion of carbon atoms located
within $M(\gg N)$ outer walls surrounding $N$ inner walls. Then,
self-contraction of the $M$ walls results in high pressure that
exerts on the $N$ walls that they encapsulate. For the sake of
analytical arguments, we simplify the $M$ walls by a continuum
elastic medium and assume the outermost boundary of the $N$-walled
tube to be in contact with the inner surface of the medium
throughout the shrinking process (see Figs.~\ref{fig_1}(b) and
(c)). As a consequence, various corrugation patterns arise in the
cross section of the $N$-walled nanotube. Only a few examples of
the corrugation patterns are shown in Fig.~\ref{fig_1}(d) together
with the associated mode index $n$ representing the wave number in
the circumferential direction.

It is noteworthy that the mechanical \cite{Xu,Yang2009} and
structural \cite{Guo} consequences of irradiation in few-walled
nanotubes have already been explored by atomistic simulations. 
Xu {\it et al.} numerically reproduced the irradiation-induced high
pressure within a double-walled nanotube\cite{Xu}; they
artificially removed carbon atoms from the outer wall and observed
the subsequent self-healing process that causes an extreme
pressure acting on the inner wall. A similar healing process is
assumed to occur when the outer walls of a many-walled nanotube
are eroded locally, as shown in Fig.~\ref{fig_1}(b). It should be
noted that in the existing experiment, it is possible to remove
carbon atoms from an MWNT with monolayer precision.\cite{JLi_2004}
This fact supports the validity of our hypothesis that a part of
the carbon atoms within only a desired number of outermost walls
is removed locally.

\subsection{Energy formulation}

The stable cross-sectional shape of the embedded tube is obtained by minimizing
its mechanical energy $U$ per unit axial length,\cite{NTN}
\begin{equation}
U = U_D + U_I + U_M + \Omega.
\label{eq_01}
\end{equation}
The first term $U_D = \sum_{i=1}^N U_D^{(i)}$ with the definition
\begin{equation}
U_D^{(i)} = \frac{r_i}{2} 
\left( 
\frac{C}{1-\nu^2} \int_0^{2\pi} \epsilon_i^2 d\theta
+
D \int_0^{2\pi} \kappa_i^2 d\theta
\right)
\end{equation}
represents the deformation energy;
$\epsilon_i$ and $\kappa_i$ are, respectively,
in-plane and bending-induced strains of the $i$th wall having the radius $r_i$
and $\theta$ is a circumferential coordinate. 
The second term  $U_I = \sum_{i,j=i\pm 1} U_I^{(i,j)}$ 
in Eq.~(\ref{eq_01}) with
$U_I^{(i,j)} = (c_{ij}r_i/2) \int_0^{2\pi} (u_i - u_j)^2 d\theta$
accounts for the van der Waals (vdW) interaction energy of all adjacent
pairs of walls. 
Here, $u_i$ is the radial displacement of the $i$th wall,
and the coefficients $c_{ij}$ are derived through a harmonic approximation
of the vdW force \cite{WBLu_APL2009}
accociated with the vdW potential $V(r) = 4\alpha [(\beta/r)^{12} - (\beta/r)^6]$
with \cite{Girifalco_PRB2000}
$\alpha=2.39$ meV and $\beta$=0.3415 nm.
The final term $\Omega$ in Eq.~(\ref{eq_01}) is
the negative of the work done by $p$ during cross-sectional deformation. 
It can be proved that\cite{NTN} all the three terms
are functions of $u_i(p,\theta)$
and the circumferential displacement $v_i(p,\theta)$ 
of the $i$th wall under $p$.

The remaining term $U_M$ in Eq.~(\ref{eq_01}) is the elastic
energy of the surrounding medium. To derive it, we assume that the
medium is homogeneous and isotropic with Young's modulus 
$E_M = 100$ GPa (that corresponds to the modulus of amorphous carbon\cite{Okada_2006,Champi_2008}) 
and Poisson's ratio $\nu_M=\nu$.
In polar coordinates, the
radial and circumferential components of normal stress in the
medium are denoted by $\sigma_r$ and $\sigma_{\theta}$,
respectively, and the shear stress is denoted by $\tau_{r\theta}$;
all the three quantities are functions of $r,\theta$. Then, $U_M$
is determined by $\sigma_r$ and $\tau_{r\theta}$ at $r=r_N$ as
\begin{eqnarray}
U_M \!\!&=&\!\! U_M^{(0)} + \Delta U_M^{(n)}, \\
U_M^{(0)}
\!\!&=&\!\! \frac{r_N}{2} \int_0^{2\pi} \!\!
\left. \sigma_r^{(0)} \right|_{r = r_N} u_N^{(0)} d\theta, \label{eq_um0} \\
\Delta U_M^{(n)}
\!\!&=&\!\!
\frac{r_N}{2} \int_0^{2\pi} \!\!
\left(\!\!
\left. \sigma_r^{(n)} \right|_{r = r_N} \delta u_N
+
\left. \tau_{r\theta}^{(n)} \right|_{r = r_N} \delta v_N
\!\!\right) d\theta, \qquad \label{eq_um}
\end{eqnarray}
where $\delta u_N$ and $\delta v_N$ describe the corrugation
amplitudes of the outermost wall of the embedded MWNT; see
Eq.~(\ref{eq_fourier}). The superscripts $(0)$ and $(n)$ indicate
that the quantities correspond to a uniform contraction and radial
corrugation, respectively. 
In plane words, $U_M^{(0)}$ represents the
energy required for uniform radial contraction of the MWNT
remaining in contact with the medium, and $\Delta U_M$ does for
radial corrugation with the mode index $n$. Details of the
derivation of $U_M$ are presented in Appendix A.

\subsection{Corrugation mode analysis}

Our objectives are to determine: (i) the optimal displacements $u_i$
and $v_i$ that minimize $U$ under a given value of $p$, and (ii)
the critical pressure $p_c$ above which the circular cross section
of an MWNT is elastically deformed into a non-circular one. These
are accomplished by the decomposition $u_i(p,\theta) = u_i^{(0)}
(p) + \delta u_i(\theta)$, where $u_i^{(0)} (p) \propto p$
describes a uniform radial contraction at $p<p_c$ and $\delta
u_i(\theta)$ describes a deformed (non-circular) cross section
observed immediately above $p_c$. Similarly, we can write
$v_i(p,\theta) = \delta v_i(\theta)$, because $v_i^{(0)}(p)\equiv
0$ at $p<p_c$. Applying the variation method to $U$ followed by
the Fourier series expansions
\begin{equation}
\delta u_i(\theta) \!\!=\!\! \sum_{n = 1}^{\infty} \delta \bar{\mu}_i(n) \cos n\theta,
\;\;
\delta v_i(\theta) \!\!=\!\! \sum_{n = 1}^{\infty} \delta \bar{\nu}_i(n) \sin n\theta,
\label{eq_fourier}
\end{equation}
we obtain the matrix equation\cite{NTN} $\bm{M} \bm{u} = \bm{0}$;
the vector $\bm{u}$ consists of $\delta \bar{\mu}_i(n)$ and
$\delta \bar{\nu}_i(n)$ with all possible $i$ and $n$ and the
matrix $\bm{M}$ involves one variable $p$ and other material
parameters. Finally, we solve the secular equation ${\rm
det}(\bm{M}) = 0$ with respect to $p$ to obtain a sequence of
discrete values of $p$ among which the minimum one serves as the
critical pressure $p_c$. Immediately above $p_c$, the circular
cross section of MWNTs becomes radially deformed, as described by
\begin{eqnarray}
u_i(\theta)
&=& u_i^{(0)}(p_c) + \delta \bar{\mu}_i(n) \cos n\theta, \label{eq_ui} \\
v_i(\theta)
&=& \delta \bar{\nu}_i(n) \sin n\theta,
\end{eqnarray}
where the value of $n$ is uniquely determined by the one-to-one
relationship between $n$ and $p_c$.

We restrict our attention to the elastic limit ({\it i.e.,} $\delta \mu_i$ and $\delta \nu_i$
be infinitesimally small);
hence, neither delamination nor strong stacking effects are considered
in the subsequent arguments,
although these effects may change the corrugation patterns..
%

\begin{figure}[ttt]
\includegraphics[width=6.0cm]{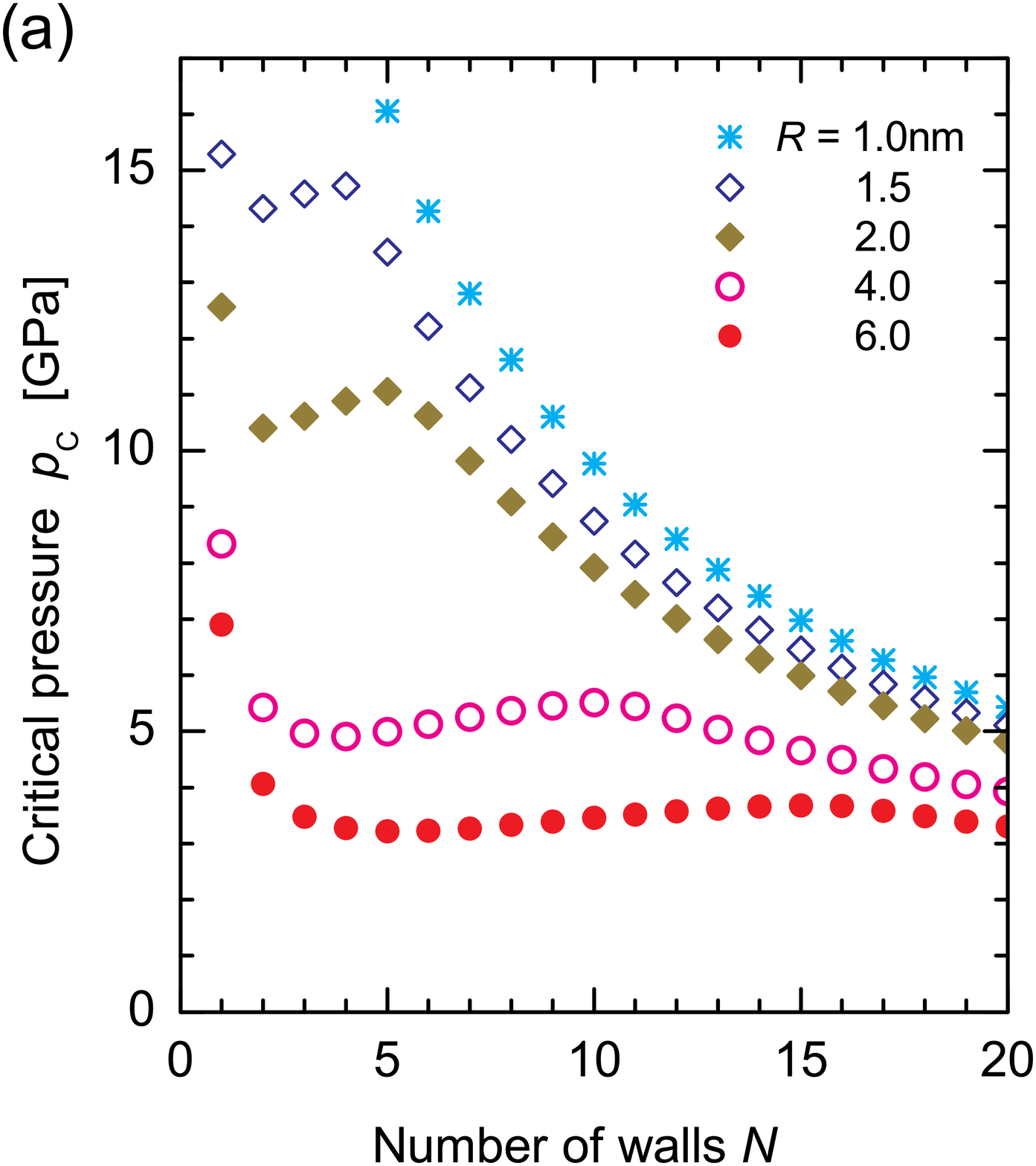}
\includegraphics[width=7.8cm]{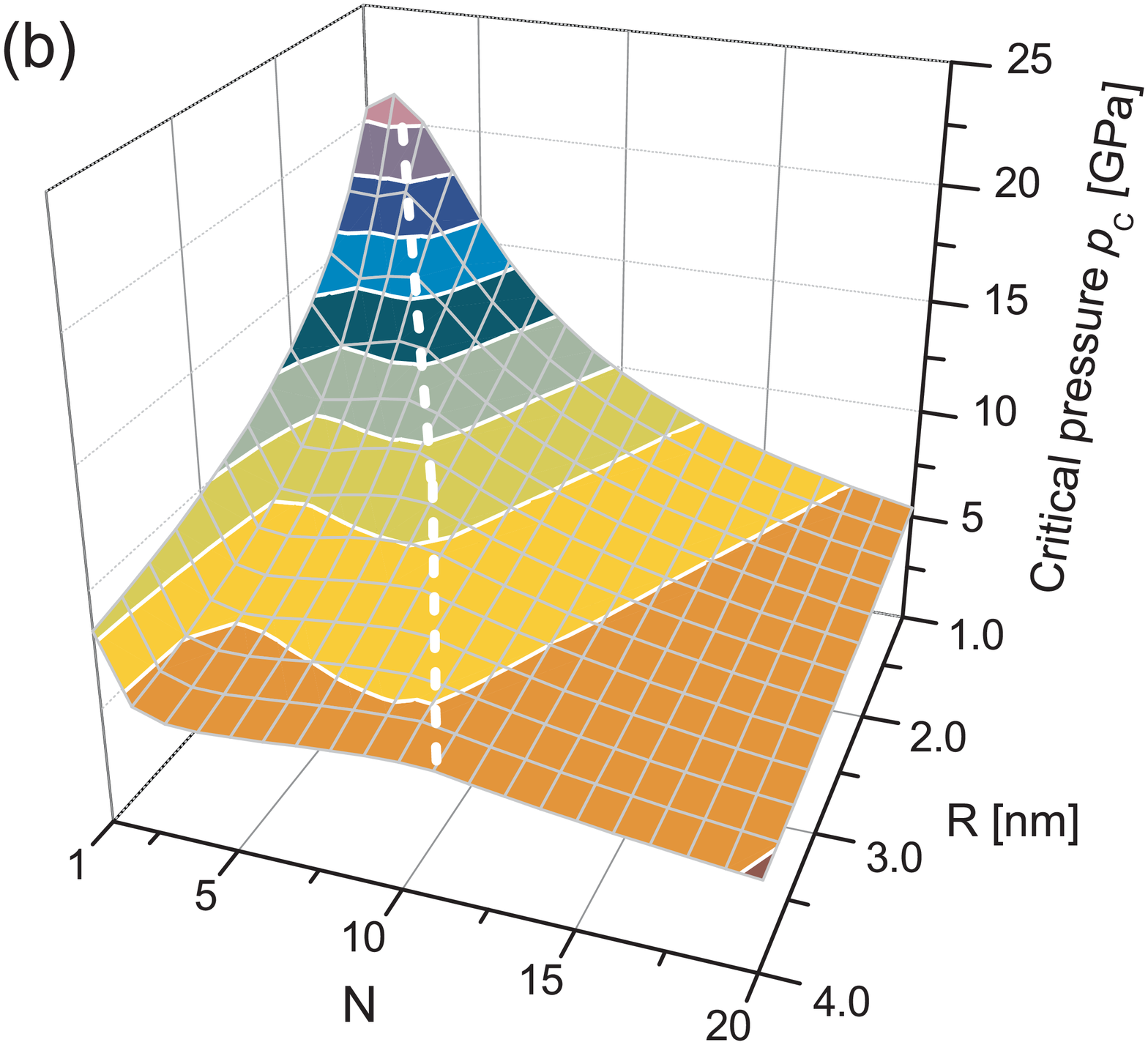}
\caption{(color online) (a) Critical pressure curve $p_c(N)$ as a function of the
total number of concentric walls $N$ contained in a MWNT. 
The innermost tube radius $R$ ranges from 1.0 nm to 6.0 nm
as indicated.
(b) Three-dimensional plot of
$p_c(R,N)$ on the $R$-$N$ plane. A ridge line extending from the
top to the skirt of the $p_c$-surface,
depicted by a bright dashed curve, corresponds to the phase
boundary that separates the inward-deformation phase from the
outward-deformation phase (see text for definitions of the two phases).} \label{fig_2}
\end{figure}

\section{Results}

\subsection{Cross-sectional view}

Figure \ref{fig_1}(d) illustrates a cross-sectional view of
typical deformation modes: the two left-hand-side panels show
``inward-deformation" modes with radial-corrugation mode indexes
$n=4$ and $5$, and the right-hand-side panel shows an
``outward-deformation" mode with $n=9$. In the inward-deformation
mode, the innermost walls exhibit significant corrugation
amplitudes as compared to the outside walls. Conversely, in the
outward-deformation mode, the innermost wall maintains its
circular shape. Which class of modes is observed immediately above
$p_c$ depends on the values of the innermost tube radius $R (\equiv r_1)$ and
$N$ under consideration. As shown below, larger $R$ and smaller
$N$ favor the inward mode with larger $n$.

\subsection{Critical pressure for radial corrugations}

Figure \ref{fig_2} (a) shows the $N$-dependence of $p_c$
for various conditions of $R$.
For all $R$s, $p_c$ exhibits
two shallow peaks (one upward and one downward) whose positions
shift to larger $N$ with an increase in $R$. 
One might expect that
a larger $N$ leads to a larger $p_c$, because an increase in the
concentric walls would enhance the radial rigidity. This
conjecture holds for MWNTs with intermediate values of $N$ between
the two shallow peaks; for instance, $p_c$ for 
$R=4.0$ nm increases slowly with $N$ between $N=4$ and $10$.
However, to the right of the upward peak, $p_c$ decreases monotonically with $N$,
in contrast to the conjecture above. 
Such decay of $p_c$ at large $N$ arises from the mechanical
instability of outside walls whose radii grow with $N$,
implying the occurrence of outward-deformation modes at large $N$.

It also follows from Fig.~\ref{fig_2}(a) that
a large portion of $p_c$'s data lies on the order of several GPa,
though they occasionally exceed ten GPa or more at certain limited conditions.
In the latter conditions, the continuum shell approximation
may break down due to interwall sp$^3$/sp$^2$ hybridization bonds
generated by pressure.
In fact, earlier MD simulations \cite{Guo}
suggested that radial pressure of approximately 40 GPa (or 30 GPa)
destroys the tubular structure of an encapsulated SWNT
having the radius $\sim 0.7$ nm ($\sim$ 1.2 nm)
through the formation of hybrid sp$^2$/sp$^3$ self linkages inside the tube.
It is thus conjectured that similar hybrid bonds between neighboring walls
in MWNTs for $R\ge 1.0$ nm
occurs above a threshold pressure comparable to (or maybe less than)
tens GPa.
Precise determination of the threshold requires atomistic calculations;
hence we proceed arguments bearing in mind that
the obtained results are physically relevant only when $p_c$ is lower than
the threshold (more or less tens GPa) that remains to be determined.

\begin{figure*}[ttt]
\includegraphics[width=7.8cm]{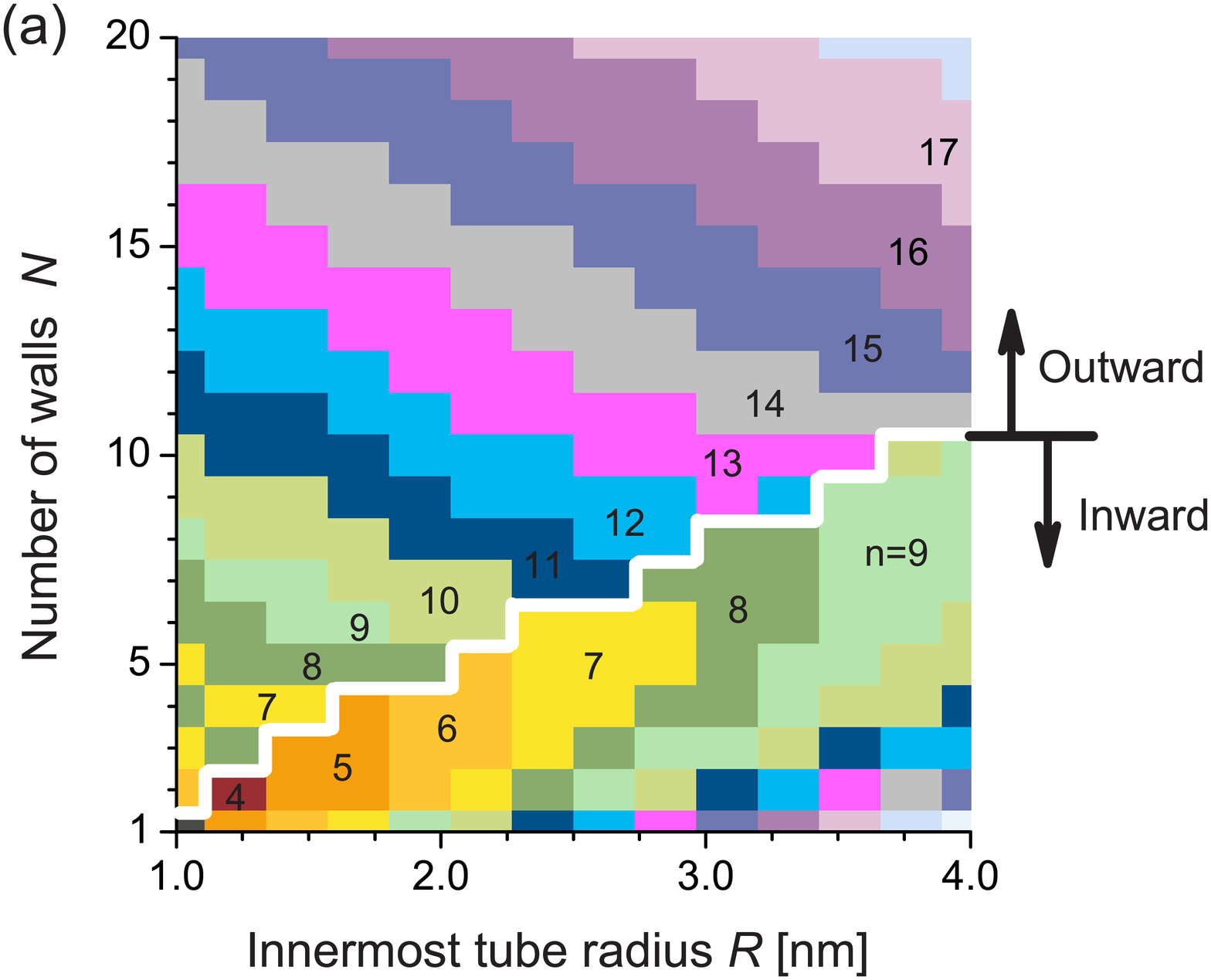}
\includegraphics[width=8.4cm]{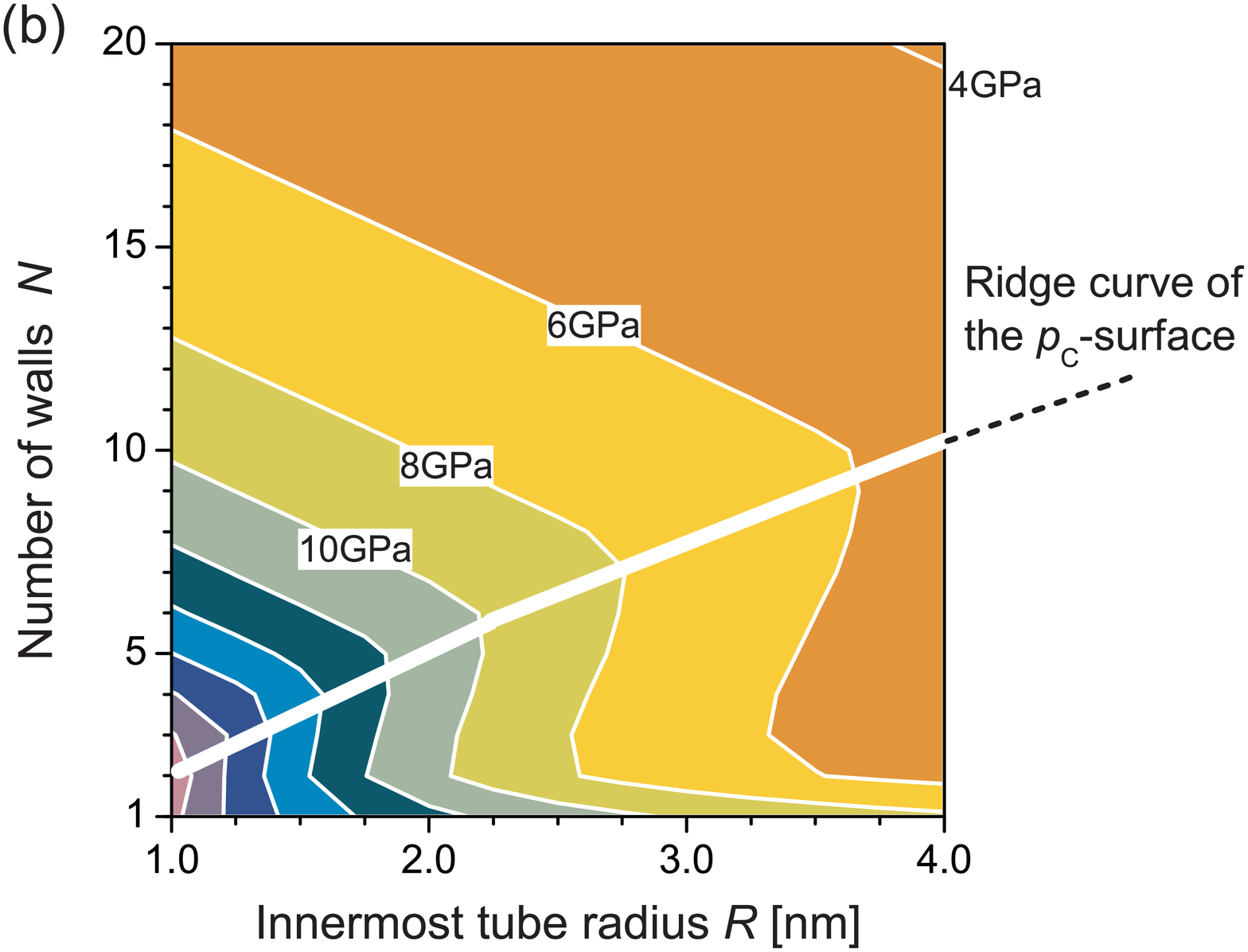}
\caption{(color online) (a) Phase diagram of the radial corrugation for embedded
MWNTs. A slanted solid line represents the phase boundary between
the inward-deformation phase (below the line) and the outward
phase (above). (b) Contour plot of the critical pressure
$p_c(R,N)$ on the $R$-$N$ plane. A solid line represents the ridge
line appearing in the three-dimensional $p_c$-surface (see
Fig.~\ref{fig_2}(b)). } \label{fig_3}
\end{figure*}

Figure \ref{fig_2}(b) shows the three-dimensional plot $p_c$ on
the $R$-$N$ plane. The plot exhibits a ridge line that extends
from the top at $(R,N) = (1.0, 2)$ to the skirt 
at $(R,N) = (4.0, 10)$ of the $p_c$-surface. The ridge line corresponds to a
phase boundary that separates the inward- and outward-deformation
phases. In fact, an inward-deformation mode occurs in the region
to the left of the ridge line, as we will find in the phase diagram in
Fig.~\ref{fig_3} (a).

\subsection{Phase diagram}

Figure \ref{fig_3}(a) shows a phase diagram of the radial
corrugation modes in MWNTs observed above $p_c$. 
A stepwise bright line represents the phase boundary between the inward-deformation
phase (below the line) and the outward phase (above). Figure
\ref{fig_3}(b) shows a contour plot of $p_c(R,N)$ that evidences a
strong correlation between the ridge line of $p_c(R,N)$ 
({\it i.e.,} the solid slanted line in Fig.~\ref{fig_3}(b))
and the phase boundary shown in Fig.~\ref{fig_3}(a).

A salient feature of Fig.~\ref{fig_3}(a) is the absence of an
elliptic deformation phase ($n=2$) within the ranges of $R$ and
$N$ we have considered. The absence of the $n=2$ mode is in
contrast with the cases of MWNTs and SWNTs under hydrostatic
pressure; in fact, the $n=2$ mode in the latter two cases is
the primary mode observed in a large domain of the $R$-$N$ space.
Furthermore, we have obtained a wide variety of corrugation modes
for various values of $R$ and $N$, where the variation of $n$ 
is systematic with the changes in $R$ and $N$. In the
inward-deformation phase, for instance, larger $R$ and smaller $N$
favor corrugation modes with large $n$. Contrariwise, in the
outward phase, smaller $R$ and larger $N$ favor corrugation modes
with large $n$.

\subsection{Corrugation amplitudes}

\begin{figure}[ttt]
\includegraphics[width=5.4cm]{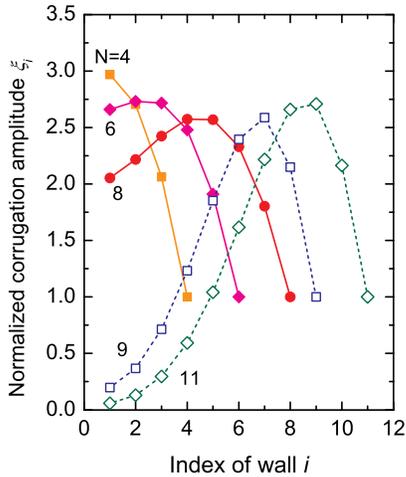}
\caption{
(color online) Normalized corrugation amplitude
$\xi_i \equiv |\delta \bar{\mu}_i/\delta \bar{\mu}_N|$
of each $i$-th concentric wall of
$N$-walled nanotubes with $R = 6.0$ nm.
Inward- (outward-) deformation modes are characterized by the value of
$\xi_1$ larger (smaller) than $1.0$.}
\label{fig_4}
\end{figure}

\begin{figure}[ttt]
\includegraphics[width=4.2cm]{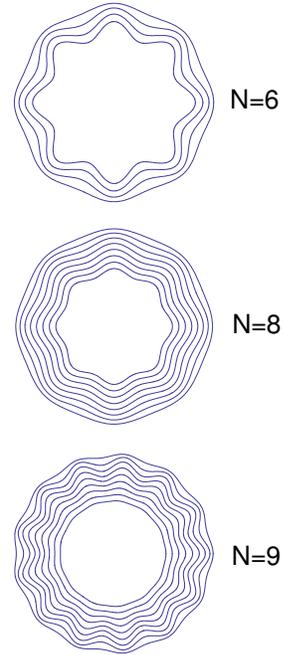}
\caption{
(color online) Cross-sectional view of the inward- ($N=6,8$)
and outward- ($N=9$) 
corrugation modes.}
\label{fig_5}
\end{figure}

The contrasting difference in corrugation amplitude distributions
between the inward and outward phases
is quantified by plotting the deformation amplitudes $\delta \bar{\mu}_i$
introduced in Eq.~(\ref{eq_ui}).
Figure \ref{fig_4} shows
the normalized deformation amplitudes, $\xi_i \equiv |\delta \bar{\mu}_i/\delta \bar{\mu}_N|$,
of individual concentric walls for $N$-walled nanotubes with different $N$
and 
$R=6.0$ nm 
be fixed.
It follows from the figure that the value of $\xi_1$ suddenly drops off
from $\xi_1 > 1.0$ to $\xi_1 < 1.0$
across the phase boundary
({\it i.e.}, from $N=8$ to $9$ with $R=6.0$ nm being fixed).
To understand this visually, we show in Fig.~\ref{fig_5}
the sequential variation of the corrugation amplitude distribution
in the cross section near the phase boundary,
In the inward mode ($N\le 8$),
the restoring force exerted by the surrounding medium is relatively strong so that
it tends to prevent radial deformation of outer walls;
this is why inner walls exhibit significant corrugation amplitudes
to lower the energy $U$ of the system.
By increasing $N$, such restoring force effects become ineffective
to support the radial instability of outer walls,
as a result of which the system falls into the outward phase for $N\ge 9$.
A similar scenario applies if we fix $N$ and modulate $R$
in the vicinity of the phase boundary.

\section{Discussions}

The present results are based on the approximation that
the radially shrinking part of an MWNT under electron-beam irradiation
is mapped onto a continuum elastic medium with
homogeneous and isotropic elasticity of the modulus $E_M= 100$ GPa.
The validity of the homogeneity and isotropy hypothesis
depends on the following two effects of irradiation on
the mechanical stiffness of MWNTs:
irradiation reduces the axial stiffness because it creates vacancies,\cite{Sammalkorpi2004,Pigno2007}
and simultaneously,
it enhances the radial stiffness owing to the production
of covalent bonding between adjacent walls.\cite{Krasheninnikov2007}
A quantitative examination of the degree to which this hypothesis holds requires
elaborated measurements or large-scale atomistic simulations,
and therefore, we have no available data for the proof.
Instead, it is possible to generalize the theoretical method
by considering the possible anisotropic elasticity in the medium.
Our preliminary calculations showed that moderate anisotropy causes
little modification in the phase diagram.
More detailed results will be shown elsewhere.

We have also performed corrugation analyses by imposing other values of $E_M$ than $E_M=100$ GPa.
It was found that larger (smaller) values of $E_M$
result in more (less) number of corrugation modes observed in the phase diagram
with fixed ranges of $R$ and $N$ set in Fig.~\ref{fig_3}(a).
With decreasing $E_M$, the phase boundary shifts downward until $E_M= 0.1$ GPa,
below which the boundary disappears leaving the outward phase as the only possibility.
Therefore, our predictions of diverse corrugation patterns
and the two corrugation phases are possible for any choice of $E_M$
as far as $E_M > 0.1$ GPa.

From an engineering perspective,
the selectivity of the innermost wall geometry by tuning the material parameters
$R$, $N$ and $p$ may be useful in developing nanotube-based nanofluidic
\cite{Majumder2005,Noy2007,Whitby2007}
or nanoelectrochemical devices \cite{Frackowiak2001},
because both utilize the hollow cavity within the innermost tube.
A very interesting issue from the academic viewpoint
is the effect of the core tube deformation on the physical and chemical properties
of intercalated molecules confined in the hollow cavity.
It has thus far been known that various types of intercalated molecules
(diatomic gas, water, organic, and transition metal molecules, {\it etc.})
can fill the innermost hollow cavities of nanotubes \cite{Noy2007}
and exhibit intriguing behaviors
that are distinct from those of
the corresponding bulk systems.\cite{Yang2003,Maniwa2007}
These distinct behaviors originate from the fact that the intermolecular spacings
become comparable to the linear dimension of the nanoscale confining space.
Therefore, the core tube deformation that breaks
the cylindrical symmetry of the initial nanoscale compartment
will engender unique properties of intercalated molecules
that are peculiar to the constrained condition in a radially corrugated space.

Another interesting implication of our results is a pressure-driven
change in the quantum transport properties of $\pi$-electrons moving along the radially corrugated nanotube.
It has been known that mobile electrons whose motion is confined to a two-dimensional curved thin layer
behave differently from those on a conventional flat plane,
because the geometric curvature of the layer
effectively yields an electromagnetic field \cite{ShimaPRB,Ono,Taira}
that can affect low-energy excitations of the electrons.
A quantitative examination along with
sophisticated atomistic simulations \cite{Arias}
should reveal novel MWNT applications
based on radial corrugation.

\section{Conclusion}

We have theoretically shown the presence of diverse radial corrugation modes
in the cross section of MWNTs that shrink radially under irradiation.
Using the continuum elastic approximation,
we have established a phase diagram that enables
a desired corrugation pattern to be obtained by tuning
the innermost wall radius $R$ and
the total number of concentric walls $N$.
We have also found that all corrugation patterns are classified into two deformation phases
between which there exists a significant difference in the corrugation amplitude of
the innermost wall of the nanotube.
We believe that the results provide useful information for developing carbon-nanotube-based
devices that utilize the nanoscale hollow cavity within the core of concentric carbon walls.

\section*{Acknowledgement}
We acknowledge M.~Rahimi, K.~Yakubo and T.~Mikami
for stimulating discussions and helpful comments.
This study was supported by a Grant-in-Aid for Scientific Research from the MEXT, Japan.
HS is thankful for the financial support provided by the Kazima Foundation and Hokkaido Gas Co., Ltd.
SG acknowledges the support of the Spanish Ministry of Science and 
Innovation through the Juan de la Cierva program. 
MA acknowledges the support of the European Research Council 
(FP7/2007-2013)/ERC grant agreement nr 240487 and the prize
``ICREA Academia'' funded by the Generalitat de Catalunya.
A part of the numerical simulations were carried out using
the facilities of the Supercomputer Center, ISSP, University of Tokyo.

\appendix
\section{Derivation of $U_M$}

In this Appendix, we derive the deformation energy $U_M$ of an elastic medium
surrounding a MWNT.
The mechanics of an elastic medium is governed by the stress function
$\phi$ that satisfies the so-called compatibility equation\cite{Timoshenko}
\begin{equation}
\left(\pa_r^2 + r^{-1}\pa_r + r^{-2}\pa_{\theta} \right)^2 \phi(r,\theta)=0,
\label{eq_app01}
\end{equation}
where $\pa_r = \pa/\pa r$ and $\pa_{\theta} = \pa/\pa \theta$.
Once $\phi$ is obtained, we can deduce the radial and circumferential
components of the normal stress, $\sigma_r$ and $\sigma_{\theta}$, respectively,
and the shear stress $\tau_{r\theta}$ as follows.
\begin{eqnarray}
\sigma_r &=& \left( r^{-1} \pa_r + r^{-2} \pa_{\theta}^2 \right)\phi, \;\;\;
\sigma_{\theta} = \pa_r^2 \phi, \label{eq_app003} \\
\tau_{r\theta} &=& \pa_r \left( r^{-1} \pa_{\theta} \right) \phi. \label{eq_app004}
\end{eqnarray}
By definition, the strain components $\ep_r, \ep_{\theta}, \gamma_{r\theta}$ are given by
the matrix form
\begin{widetext}
\begin{equation}
\left[
\begin{array}{c}
\ep_r \\
\ep_{\theta} \\
\gamma_{r\theta}
\end{array}
\right]
=
\frac{1}{E_M}
\left[
\begin{array}{ccc}
1-\nu_M^2 & -\nu_M (1+\nu_M) & 0 \\
-\nu_M (1+\nu_M) & 1-\nu_M^2 & 0 \\
0 & 0 & 2(1+\nu_M)
\end{array}
\right]
\left[
\begin{array}{c}
\sigma_r \\
\sigma_{\theta} \\
\tau_{r\theta}
\end{array}
\right]
=
\left[
\begin{array}{cc}
\pa_r & 0 \\
r^{-1} & r^{-1} \pa_{\theta} \\
r^{-1} \pa_{\theta} & \pa_r - r^{-1}
\end{array}
\right]
\left[
\begin{array}{c}
u \\
v
\end{array}
\right],
\label{eq_app_mat}
\end{equation}
\end{widetext}
where $u = u(r,\theta)$ and $v=v(r,\theta)$ are respectively the radial and circumferential displacements
of a volume element in the host medium.

The general solution of Eq.~(\ref{eq_app01}) is given by
$\phi(r,\theta) = \sum_{n=0}^{\infty} \phi_n(r,\theta)$,
where $\phi_n(r,\theta) = f_n(r) \cos n \theta + g_n(r) \sin n \theta$
and\cite{Croll}
\begin{eqnarray}
f_0 &=& \!\! a_0 \log r \!+\! b_0 r^2 \log r \!+\! c_0 r^2 \!+\! d_0 \!+\! \alpha_0 \theta \!+\! \beta_0 r^2 \theta, \label{eq_app005} \\
f_1 &=& \!\!a_1 r^{-1} \!+\! b_1 r \log r \!+\! c_1 r^3 + d_1 r \!+\! \alpha_1 \theta, \\
f_{n} &=& \!\!a_n r^{-n} \!+\! b_n r^{2-n} \!+\! c_n r^{2+n} \!+\! d_n r^n, (n\ge 2) \label{eq_app007}
\end{eqnarray}
with similar definitions of $g_n$.
The zeroth component $\phi_0$ represents a uniform contraction of the circular cross section,
thus corresponding to the energy $U_M^{(0)}$ that we have introduced in Eq.~(\ref{eq_um0}).
The first one $\phi_1$ implies a rigid body translation that is irrelevant to our consideration.
Other components $\phi_n$ for $n\ge 2$ describe radial corrugations with mode index $n$,
thus providing the energy $\Delta U_M$ given by Eq.~(\ref{eq_um}).
In the following, we set $f_0 = a_0 \log r$ and $f_n = a_n r^{-n} + b_n r^{2-n}$
instead of Eqs.~(\ref{eq_app005}) and (\ref{eq_app007}), respectively,
in order to obtain physically relevant solutions of
$\sigma_r, \sigma_{\theta}, \tau_{r\theta}$ that decay with increasing $r$.\cite{Croll}

We now evaluate the explicit forms of $U_M^{(0)}$ and $\Delta U_M^{(n)}$.
To derive $U_M^{(0)}$, we consider a specific solution of (\ref{eq_app01}) that has
the form of $\phi = \phi_0$, and then
substitute it in Eqs.~(\ref{eq_app003}) and (\ref{eq_app004})
to obtain
$\sigma_r^{(0)} = a_0 r^{-2}$, $\sigma_{\theta}^{(0)} = -a_0 r^{-2}$,
and $\tau_{r\theta}^{(0)} = 0$.
Hence, it follows from Eq.~(\ref{eq_app_mat}) that
$u^{(0)}(r) = (1+\nu_M) a_0/(E_M r)$,
$v^{(0)} \equiv 0$.
Complete contact between the medium and the outermost wall implies $u^{(0)}(r_N) = u_N^{(0)}$,
and the elastic nature of the medium implies $\sigma_r^{(0)}|_{r=r_N} = \kappa_0 u_N^{(0)}$
with the stiffness coefficient $\kappa_0$.
Obviously, $\kappa_0$ is identified with $\sigma_r^{(0)}|_{r=r_N}$ associated with
the unit radial displacement $u_N^{(0)} \equiv 1$.
Hence, imposing an appropriate value of $a_0$, we obtain
$\kappa_0 = - E_M/[(1+\nu_M)r_N]$.
Moreover, the uniform contraction energy $U_M^{(0)}$ should be represented as
$U_M^{(0)} = (\kappa_0/2) \int_0^{2\pi} \left\{ u_N^{(0)} \right\}^2 r_N d\theta$.
Consequently, we obtain the explicit form of $U_M^{(0)}$
that depends on $E_M, \nu_M$ and $u_N^{(0)}$.

Next, we consider the energy $\Delta U_M^{(n)}$ $(n\ge 2)$ that corresponds to
the radial corrugation of the $n$th order.
We only consider the cosine term in $\phi_N$ without loss of generality,
which is based on our assumption of cosine radial displacement $\delta u_i(\theta)$
(see Eq.~(\ref{eq_fourier})).
A similar procedure to the case of $n=0$ yields
\begin{eqnarray}
\sigma_r^{(n)} \!\!\! &=& \!\!\!
\left\{ -n(n\!+\!1) a_n r^{-2} \!-\! (n\!-\!1) (n\!+\!2)  b_n \right\} \! r^{-n} \!
\cos n\theta, \nonumber \\
& & \\
\sigma_{\theta}^{(n)}
\!\!\! &=&\!\!\!
\left\{ n(n+1) a_n r^{-2} \!+\! (n\!-1)(n\!-\!2)  b_n \right\} \! r^{-n} \! \cos n\theta,
\nonumber \\
& & \\
\tau_{r\theta}^{(n)}
\!\!\! &=& \!\!\!
\left\{ -n(n+1) a_n r^{-2} \!-\! n(n\!-\!1)  b_n \right\} \! r^{-n} \! \sin n\theta,
\end{eqnarray}
leading to the results
\begin{eqnarray}
u^{(n)}(r,\theta)
&=& \frac{(1+\nu_M)}{E_M} \left[ n a_n r^{-2} \right.\nonumber \\
&+& \!\! \left.\left\{ n+2(1-2\nu_M) \right\} b_n \right] r^{1-n} \cos n \theta, \quad \\
v^{(n)}(r,\theta)
&=& \frac{(1+\nu_M)}{E_M} \left[ n a_n r^{-2} \right.\nonumber \\
&+& \!\! \left.\left\{ n-4(1-\nu_M) \right\} b_n \right] r^{1-n} \sin n \theta.
\end{eqnarray}
Parallel discussions to the $n=0$ case together with the formula
\begin{eqnarray}
\left. \sigma_r^{(n)} \right|_{r=r_N}
&=&
\left( k_{11} \delta \bar{\mu}_N + k_{12} \delta \bar{\nu}_N \right) \cos n \theta, \label{eq_sigma} \\
\left. \tau_{r\theta}^{(n)} \right|_{r=r_N}
&=&
\left( k_{21} \delta \bar{\mu}_N + k_{22} \delta \bar{\nu}_N \right) \sin n \theta,
\end{eqnarray}
lead us to attain the stiffness coefficients
$k_{11} = -\sigma_r^{(n)} |_{r=r_N, \delta \bar{\mu}_N=1, \delta \bar{\nu}_N=0}/\cos n \theta$
and
$k_{12} = -\sigma_r^{(n)} |_{r=r_N, \delta \bar{\mu}_N=0, \delta \bar{\nu}_N=1}/\cos n \theta$
as
\begin{eqnarray}
k_{11} 
&=&
\frac{2(n+1)(1-\nu_M) - 1}{(1+\nu_M)(3-4\nu_M)}E_M, \\
k_{12} 
&=&
\frac{2(n+1)(1-\nu_M) - n}{(1+\nu_M)(3-4\nu_M)}E_M, \label{eq_k12}
\end{eqnarray}
and $k_{22} = k_{11}$, $k_{21} = k_{12}$.
Substituting the results (\ref{eq_sigma})-(\ref{eq_k12}) into Eq.~(\ref{eq_um}),
we finally obtain the explicit form of $\Delta U_M^{(n)}$.

%
%


\begin{thebibliography}{99}

\bibitem{Despres1995}
J.~F.~Despres, E.~Daguerre, and K.~Lafdi, Carbon {\bf 33}, 87 (1995).
\bibitem{Iijima1996}
S.~Iijima, C.~Brabec, A.~Maiti, and J.~Bernholc,
J.~Chem.~Phys. {\bf 104}, 2089 (1996).
\bibitem{Palaci2005}
I.~Palaci, S.~Fedrigo, H.~Brune, C.~Klinke, M.~Chen, and E.~Riedo,
Phys.~Rev.~Lett. {\bf 94}, 175502 (2005).
\bibitem{J_Tang2000}
J.~Tang, L.~C.~Qin, T.~Sasaki, M.~Yudasaka, A.~Matsushita, and S.~Iijima,
Phys.~Rev.~Lett. {\bf 85}, 1887 (2000).
\bibitem{Peters2000}
M.~J.~Peters, L.~E.~McNeil, J.~P.~Lu, and D.~Kahn,
Phys.~Rev.~B {\bf 61}, 5939 (2000).
\bibitem{Elliott2004}
J.~A.~Elliott, J.~K.~W.~Sandler, A.~H.~Windle, R.~J.~Young, and M.~S.~P.~Shaffer,
Phys.~Rev.~Lett. {\bf 92}, 095501 (2004).
\bibitem{Tangney2005}
P.~Tangney, R.~B.~Capaz, C.~D.~Spataru, M.~L.~Cohen, and S.~G.~Louie,
Nano Lett. {\bf 5}, 2268 (2005).
\bibitem{Gadagkar2006}
V~Gadagkar, P.~K.~Maiti, Y.~Lansac, A.~Jagota, and A.~K.~Sood,
Phys. Rev. B {\bf 73}, 085402 (2006).
\bibitem{Zhang2006}
S.~Zhang, R.~Khare, T.~Belytschko, K.~J.~Hsia, S.~L.~Mielke, and G.~C.~Schatz,
Phys.~Rev.~B {\bf 73}, 075423 (2006).
\bibitem{Reich2002}
S.~Reich, C.~Thomsen, and P.~Ordejon, Phys.~Rev.~B {\bf 65}, 153407 (2002).
\bibitem{Chan2003}
S.~P.~Chan, W.~L.~Yim, X.~G.~Gong, and Z.~F.~Liu, Phys.~Rev.~B {\bf 68}, 075404 (2003).
\bibitem{Barboza_2008}
A.~P.~M.~Barboza, A.~P.~Gomes, B.~S.~Archanjo, P.~T.~Araujo, A.~Jorio, A.~S.~Ferlauto,
M.~S.~C.~Mazzoni, H.~Chacham, and B.~R.~A.~Neves,
Phys.~Rev.~Lett. {\bf 100}, 256804 (2008).
\bibitem{Barboza_2009} 
A.~P.~M.~Barboza, H.~Chacham, and B.~R.~A.~Neves, Phys.~Rev.~Lett. {\bf 102}, 025501 (2009).
\bibitem{Park1999}
C.~J.~Park, Y.~H.~Kim, and K.~J.~Chang, Phys.~Rev.~B {\bf 60}, 10656 (1999).
\bibitem{Mazzoni2000}
M.~S.~C.~Mazzoni and H.~Chacham, Appl.~Phys.~Lett. {\bf 76}, 1561 (2000).
\bibitem{Gomez2006}
C.~G\'omez-Navarro, J.~J.~S\'aenz, and J.~G\'omez-Herrero, Phys.~Rev.~Lett. {\bf 96}, 076803 (2006).
\bibitem{Lebedkin2006}
S.~Lebedkin, K.~Arnold, O.~Kiowski, F.~Hennrich, and M.~M.~Kappes,
Phys.~Rev.~B {\bf 73}, 094109 (2006).
\bibitem{Yakobson1996}
B.~I.~Yakobson, C.~J.~Brabec, and J.~Bernholc, Phys.~Rev.~Lett. {\bf 76}, 2511 (1996).
\bibitem{DYSun2004}
D.~Y.~Sun, D.~J.~Shu, M.~Ji, F.~Liu, M.~Wang, and X.~G.~Gong,
Phys.~Rev.~B {\bf 70}, 165417 (2004).
\bibitem{Brush_1975}
D.~O.~Brush and B.~O.~Almroth, {\it Buckling of bars, plates, and shells} (McGraw-Hill, 1975).
\bibitem{NTN}
H.~Shima and M.~Sato, Nanotechnology {\bf 19}, 495705 (2008);
phys.~stat.~sol.~(a) {\bf 206}, 2228 (2009).
\bibitem{Allen_1969}
H.~G.~Allen, {\it Analysis and Design of Structural Sandwich Panels}, (Pergamon Press, Oxford, 1969).
\bibitem{Volynskii_2000}
A.~L.~Volynskii, S.~Bazhenov, O.~V.~Lebedeva, and N.~F.~Bakeev, 
J.~Mater.~Sci. {\bf 35}, 547 (2000).
\bibitem{Majumder2005}
M.~Majumder, N.~Chopra, R.~Andrews, and B.~J.~Hinds, Nature (London) {\bf 438}, 44 (2005).
\bibitem{Noy2007}
A.~Noy, H.~G.~Park, F.~Fornasiero, J.~K.~Holt, C.~P.~Grigoropoulos, and O.~Bakajin, Nano Today {\bf 2}, 22 (2007).
\bibitem{Whitby2007}
M.~Whitby and N.~Quirke, Nature Nanotech. {\bf 2}, 87 (2007).
\bibitem{Frackowiak2001}
E.~Frackowiak and F.~Beguin, Carbon {\bf 39}, 937 (2001); {\it ibid.} {\bf 40}, 1775 (2002).
\bibitem{Banhart1996}
F.~Banhart and P.~M.~Ajayan, Nature {\bf 382}, 433 (1996);
Adv.~Mater. {\bf 9}, 261 (1997).
\bibitem{Ajayan1998}
P.~M.~Ajayan, V.~Ravikumar, and J.~C.~Charlier, Phys.~Rev.~Lett. {\bf 81}, 1437 (1998).
\bibitem{Banhart2005}
F.~Banhart, J.~X.~Li, and A.~V.~Krasheninnikov, Phys.~Rev.~B {\bf 71}, 241408(R) (2005).
\bibitem{LSun2006}
L.~Sun, F.~Banhart, A.~V.~Krasheninnikov, J.~A.~Rodr\'iguez-Manzo, M.~Terrons, and P.~M.~Ajayan,
Science {\bf 312}, 1199 (2006).
\bibitem{Krasheninnikov2007}
A.~V.~Krasheninnikov and  F.~Banhart, Nature Mat. {\bf 6}, 723 (2007).
\bibitem{Ding2007}
F.~Ding, K.~Jiao, Y.~Lin, and B.~I.~Yakobson,
Nano Lett. {\bf 7}, 681 (2007);
F.~Ding, K.~Jiao, M.~Wu, and B.~I.~Yakobson,
Phys.~Rev.~Lett. {\bf 98}, 075503 (2007);
J.~Y.~Huang, F.~Ding, K.~Jiao, and B.~I.~Yakobson,
{\it ibid.,} Phys.~Rev.~Lett. {\bf 99}, 175503 (2007).
\bibitem{Ru}
C.~Q.~Ru, Phys. Rev. B {\bf 62}, 16962 (2000).
\bibitem{Pantano2004}
A.~Pantano, D.~M.~Parks, and M.~C.~Boyce, J.~Mech.~Phys.~Solid. {\bf 52}, 789 (2004).
\bibitem{Huang}
Y.~Huang, J.~Wu, and K.~C.~Hwang, Phys.~Rev.~B {\bf 74}, 245413 (2006).
\bibitem{Kudin_PRB2001}
K.~N.~Kudin, G.~E.~Scuseria, and B.~I.~Yakobson,
Phys.~Rev.~B {\bf 64}, 235406 (2001).
\bibitem{Xu}
Z.~Xu, L.~Wang, and Q.~Zheng, Small {\bf 4}, 733 (2008).
\bibitem{Yang2009}
S.~H.~Yang, L.~L.~Feng and F.~Feng, J.~Phys.~D: Appl.~Phys. {\bf 42}, 055414 (2009).
\bibitem{Guo}
Y.~Guo and W.~Guo, Phys.~Rev.~B {\bf 76}, 045404 (2007).
\bibitem{JLi_2004}
J.~Li and F.~Banhart, Nano Lett. {\bf 4}, 1143 (2004).
\bibitem{WBLu_APL2009}
W.~B.~Lu, B.~Liu, J.~Wu, J.~Xiao, K.~C.~Hwang, S.~Y.~Fu, and Y. Huang,
Appl.~Phys.~Lett. {\bf 94}, 101917 (2009).
\bibitem{Girifalco_PRB2000}
L.~A.~Girifalco, M.~Hodak, and R.~S.~Lee, Phys.~Rev.~B {\bf 62}, 13104 (2000).
\bibitem{Okada_2006}
S.~Okada, T.~Mukawa, R.~Kobayashi, M.~Ishida, Y.~Ochiai,
T.~Kaito, S.~Matsui, and J.~Fujita,
Jpn.~J.~Appl.~Phys. {\bf 45}, 5556 (2006).
\bibitem{Champi_2008}
A.~Champi, A.~S.~Ferlauto, F.~Alvarez, S.~R.~P.~Silva, and F.~C.~Marques,
Diamond Relat.~Mater. {\bf 17}, 1850 (2008).
\bibitem{Sammalkorpi2004}
M.~Sammalkorpi, A.~Krasheninnikov, A.~Kuronen, K.~Nordlund, and K.~Kaski
Phys.~Rev.~B {\bf 70}, 245416 (2004).
\bibitem{Pigno2007}
N.~M.~Pugno, Appl.~Phys.~Lett. {\bf 90}, 043106 (2007).
\bibitem{Yang2003}
C.~K.~Yang, J.~Zhao, and J.~P.~Lu, Phys.~Rev.~Lett. {\bf 90} (2003) 257203.
\bibitem{Maniwa2007}
Y.~Maniwa, K.~Matsuda, H.~Kyakuno, S.~Ogasawara, T.~Hibi, H.~Kadowaki, S.~Suzuki, Y.~Achiba, and H.~Kataura,
Nature Mater. {\bf 6}, 135 (2007).
\bibitem{ShimaPRB}
H.~Shima, H.~Yoshioka and J.~Onoe,
Phys.~Rev.~B {\bf 79}, 201401(R) (2009); Physica E {\bf 42}, 1151 (2010).
\bibitem{Ono}
S.~Ono and H.~Shima, Phys.~Rev.~B {\bf 79}, 235407 (2009); Physica E {\bf 42}, 1224 (2010).
\bibitem{Taira}
H.~Taira and H.~Shima, J.~Phys.:~Condens.~Mat. {\bf 22}, 075301 (2010);
{\it ibid.}, {\bf 22}, 245302 (2010).
\bibitem{Arias}
I.~Arias and M.~Arroyo, Phys.~Rev.~Lett. {\bf 100}, 085503 (2008);
M.~Arroyo and I.~Arias, J.~Mech.~Phys.~Solids {\bf 56}, 1224 (2008).
\bibitem{Timoshenko}
S.~R.~Timoshenko and J.~N. Goodier, {\it Theory of elasticity}, 3rd Ed. (McGraw-Hill, 1970).
\bibitem{Croll}
J.~G.~A.~Croll, J.~Eng.~Mech. {\bf 127}, 333 (2001).

\end{thebibliography}
\end{document}